\documentclass[draft]{agujournal2019}
\usepackage{url} %this package should fix any errors with URLs in refs.
\usepackage{lineno}
\usepackage[inline]{trackchanges} %for better track changes. finalnew option will compile document with changes incorporated.
\usepackage{soul}
\usepackage{amssymb,amsmath}
\usepackage[autostyle]{csquotes}
\draftfalse

\journalname{Geophysical Research Letters}

\begin{document}

%% ------------------------------------------------------------------------ %%
%  Title
%
% (A title should be specific, informative, and brief. Use
% abbreviations only if they are defined in the abstract. Titles that
% start with general keywords then specific terms are optimized in
% searches)
%
%% ------------------------------------------------------------------------ %%

\title{
A youthful Titan implied by improved impact simulations
}
%Geophysical Research Letters have a maximum length of 12 publication units.
%For most journals, Research Articles are allowed to be up to 25 publication units (PU), where 1 PU is 500 words or 1 display element (figure or table). 

%% ------------------------------------------------------------------------ %%
%
%  AUTHORS AND AFFILIATIONS
%
%% ------------------------------------------------------------------------ %%

% Authors are individuals who have significantly contributed to the
% research and preparation of the article. Group authors are allowed, if
% each author in the group is separately identified in an appendix.)

% List authors by first name or initial followed by last name and
% separated by commas. Use \affil{} to number affiliations, and
% \thanks{} for author notes.
% Additional author notes should be indicated with \thanks{} (for
% example, for current addresses).

\authors{S. Wakita\affil{1,2}, B. C. Johnson\affil{1,3}, J. M. Soderblom\affil{2}, C. D. Neish\affil{4,5}}

\affiliation{1}{Department of Earth, Atmospheric, and Planetary Sciences, Purdue University, West Lafayette, IN, USA}
\affiliation{2}{Department of Earth, Atmospheric and Planetary Sciences, Massachusetts Institute of Technology, Cambridge, MA, USA}
\affiliation{3}{Department of Physics and Astronomy, Purdue University, West Lafayette, IN, USA}
\affiliation{4}{The Planetary Science Institute, Tucson, AZ, USA}
\affiliation{5}{Department of Earth Sciences, The University of Western Ontario, London, ON, Canada}

%% Corresponding Author:
% (include name and email addresses of the corresponding author.  More
% than one corresponding author is allowed in this LaTeX file and for
% publication; but only one corresponding author is allowed in our
% editorial system.)

\correspondingauthor{Shigeru Wakita}{swakita@purdue.edu}

%% Keypoints, final entry on title page.

%  List up to three key points (at least one is required)
%  Key Points summarize the main points and conclusions of the article
%  Each must be 140 characters or fewer with no special characters or punctuation and must be complete sentences

\begin{keypoints}
\item To produce new impact crater scaling laws for Titan, we simulate impacts with and without a methane-clathrate layer.
\item Using these new scaling laws, we find that Titan's crater retention age is 300--340 Myr, assuming a surface methane-clathrate layer.
\item Titan's surface retention age is extremely young compared to other icy satellites, suggesting ongoing geologic activity.
\end{keypoints}

%% ------------------------------------------------------------------------ %%
%
%  ABSTRACT and PLAIN LANGUAGE SUMMARY
%
% A good Abstract will begin with a short description of the problem
% being addressed, briefly describe the new data or analyses, then
% briefly states the main conclusion(s) and how they are supported and
% uncertainties.

% The Plain Language Summary should be written for a broad audience,
% including journalists and the science-interested public, that will not have 
% a background in your field.
%
% A Plain Language Summary is required in GRL, JGR: Planets, JGR: Biogeosciences,
% JGR: Oceans, G-Cubed, Reviews of Geophysics, and JAMES.
% see http://sharingscience.agu.org/creating-plain-language-summary/)
%
%% ------------------------------------------------------------------------ %%

%% \begin{abstract} starts the second page

%For Geophysical Research Letters (GRL), your abstract should be less than 150 words.
\begin{abstract}
The small number of impact craters found on Titan suggests that its surface is relatively young. 
Previous work estimated its surface age to be between 200 and 1000 Myr. 
This estimate, however, is based on crater scaling laws for water and sand, which are not representative of the composition of Titan's icy surface. 
Titan's surface is likely composed of water ice, methane clathrates, or a combination of both. 
Here, we perform impact simulations for impactors of various sizes that strike an icy target with a 0--15 km thick methane clathrate cap layer. 
We derive new crater scaling laws based on our numerical results, and find that Titan's surface age is 300--340 Myr, assuming heliocentric impactors and surface clathrates.
This age, which represents the crater retention age, indicates a relatively youthful surface, suggesting that active endogenic and/or exogenic processes have recently reshaped Titan's surface.
\end{abstract}

%A PLS should be no longer than 200 words and should be free of jargon, acronyms, equations, and any technical information that would be unknown to people from outside your scientific discipline.
\section*{Plain Language Summary}
The largest Saturnian moon, Titan, has only 90 potential impact craters below its thick atmosphere. 
This number of craters is smaller than that of other similarly sized icy moons, indicating that Titan's surface is geologically young.
Previous work estimated its age as 200--1000 million years based on crater formation in water or sand, however, this approach might not be suitable for Titan. 
Atmospheric methane leads to methane rain that can interact with the water ice crust and form %methane clathrate. 
a crystal structure of ice that traps methane molecules. 
This is known as methane clathrate, that has different properties from pure water-ice.
When methane clathrate covers Titan's surface, it affects crater formation. 
Here, we investigate the formation of impact craters in methane clathrate to revise Titan's surface age. 
Our simulations demonstrate that craters formed in a methane-clathrate layer have larger diameters than those formed in a pure water ice crust. 
We find that Titan's surface age is 300--340 million years, assuming a surface clathrate, similar to the younger end of the previous estimate. 
This suggests that Titan's surface is actively being degraded over time, or a significant event erased Titan's landscape hundreds of millions of years ago.

%% ------------------------------------------------------------------------ %%
%
%  TEXT
%
%% ------------------------------------------------------------------------ %%

%%% Suggested section heads:
% \section{Introduction}
%
% The main text should start with an introduction. Except for short
% manuscripts (such as comments and replies), the text should be divided
% into sections, each with its own heading.

% Headings should be sentence fragments and do not begin with a
% lowercase letter or number. Examples of good headings are:

% \section{Materials and Methods}
% Here is text on Materials and Methods.
%
% \subsection{A descriptive heading about methods}
% More about Methods.
%
% \section{Data} (Or section title might be a descriptive heading about data)
%
% \section{Results} (Or section title might be a descriptive heading about the
% results)
%
% \section{Conclusions}

\section{Introduction} \label{sec:introduction}
With only 90 possible craters, Titan’s surface is much more sparsely cratered than most other icy satellites. 
This remains true even if you only consider those craters unaffected by disruption in Titan's thick atmosphere ($D_{\rm crater} \gtrsim 20$ km), implying that Titan has a relatively young surface.
Current studies constrain its age to be between 200 and 1000 Myr, depending on the crater scaling law used \cite{Neish:2012,Hedgepeth:2020}. 
This age, more accurately referred to as the crater retention age, is the duration for which craters on Titan remain detectable in Cassini data \cite<cf.>{Hartmann:1966} before they are eroded, buried, or replaced with new crustal material.
Constraining Titan's crater retention age (hereafter referred to as the \enquote{surface age}) provides critical information regarding its geologic history and associated landscape evolution. 
For example, the age of Titan's surface may constrain the minimum age of Titan’s methane-rich atmosphere \cite{Nixon:2012,Mandt:2012,Horst:2008,Horst:2017}, and the rates at which Titan's landscape is modified, providing insight into the formation and erosion of landforms on Titan \cite{Lorenz:2007,Moore:2011,Birch:2017a,Corlies:2017,Neish:2016,Schurmeier:2024}. 
The current constraints on Titan's surface age assume heliocentric impactors \cite{Zahnle:2003}, and crater scaling laws based on water and sand targets \cite{Artemieva:2005,Korycansky:2005}. 
Recent work \cite{Wakita:2022,Wakita:2023}, however, suggests that these crater scaling laws may be inappropriate for Titan.

Titan has a dense atmosphere dominated by nitrogen, with methane as the second most abundant molecule \cite{Kuiper:1944,Broadfoot:1981}.
This atmosphere supports a methane-based hydrological cycle, which produces methane rain \cite{Turtle:2009,Turtle:2011} and methane lakes in the polar regions \cite{Stofan:2007,Hayes:2008}. 
When methane rains and soaks into the porous water-ice crust, methane clathrates may form in the crust.
Methane clathrate is another crystal structure of water-ice, which traps methane molecules \cite{Sloan:2007}.
Such clathrates are thought to form quickly under Titan's surface temperature and pressure \cite{Choukroun:2010,Vu:2020}, and are stable at Titan’s surface conditions \cite{Tobie:2006}. 
Separately, it is possible that Titan's methane is endogenic, and degassing due to the dehydration process of Titan's core \cite{Castillo-Rogez:2010,Miller:2019} may release methane that forms clathrates in the crust as it ascends to the surface \cite{Tobie:2006,Carnahan:2022}.

Methane clathrate has unique properties that significantly influence crater formation \cite{Wakita:2023}. 
Methane clathrate is 20 times stronger than water ice at temperatures near the melting point of water ice \cite{Durham:2003}, resulting in the formation of slightly smaller craters when the temperature profile in the crust is the same \cite{Wakita:2022}.
More significantly, however, is that methane clathrate exhibits 4 times lower thermal conductivity than water ice at 263 K \cite{Sloan:2007}. 
Consequently, a methane-clathrate layer acts as an insulator, leading to a higher near-surface temperature gradient compared to a pure water-ice crust \cite{Kalousova:2020}. 
Because the temperature profile in the target affects its yield strength, it strongly influences crater morphology \cite{Bray:2014,Silber:2017,Bjonnes:2022}. 
It is essential to investigate the effect of a methane-clathrate layer on crater diameter when determining Titan's surface age. 
Here, we simulate impacts striking icy targets with and without methane-clathrate cap layers using a shock physics code. 
We derive new crater scaling laws for targets composed of pure water-ice and those with methane-clathrate cap layers, and use them to reevaluate Titan's surface age.

\section{Methods} \label{sec:methods}
We simulate impacts on Titan using the iSALE-2D shock physics code.
Based on the SALE hydrocode \cite{Amsden:1980}, iSALE has been developed to model planetary impacts and cratering \cite{Collins:2016, Wunnemann:2006}.
The code has been improved from the SALE code by including various equations of state (EOS) and strength models \cite{Collins:2004, Ivanov:1997, Melosh:1992}. 
We assume that a spherical water-ice impactor with no porosity vertically strikes a flat target at 10.5 km/s, the average impact velocity on Titan \cite{Zahnle:2003}.
Note this is slightly lower than the average impact velocity used in \citeA{Nesvorny:2023}, 11.2 km/s. 
Although our 10.5 km/s impact would form a crater that is 3 \% smaller in diameter \cite<e.g., $v_{\rm imp}^{0.5}$, see>{Johnson:2016c}, this has a negligible effect on our results.
We consider impactor diameters of 1.5 km, the smallest size that is minimally affected during atmospheric entry \cite{Artemieva:2003}, to 25 km, the estimated size of the impactor \cite{Crosta:2021} that formed the largest crater observed on Titan, Menrva \cite<400 km,>{Hedgepeth:2020}.
For the computational domain, iSALE-2D has a high-resolution zone and an extension zone. 
To save computational resources, the cell size in the high-resolution zone varies according to the impactor size: 50 m for 1.5--5 km, 100 m for 6 \& 8 km, 200 m for 10--20 km, and 250 m for 25 km.
Note that we use previous data collected for 3--4 km diameter impactors, which also used a 50 m resolution \cite{Wakita:2023}.

To explore the effect of methane clathrate on the resultant crater morphometry, we vary the thickness of the methane-clathrate layer. 
We consider 0, 5, 10, and 15 km thick methane-clathrate layers with corresponding thermal profiles that assume 1 mm ice grains \cite<see Figure \ref{fig:tmp_yld}a and>{Kalousova:2020}.
Our choice of methane-clathrate layer thickness is consistent with previous work \cite{Kalousova:2020,Wakita:2023}.
We use the steady-state thermal profiles reported by \citeA{Kalousova:2020}, which are derived from two-dimensional simulations that assume a 100 km-thick ice shell composed of conductive and convective ice and ignore tidal heating and radiogenic heating. 
To model the strength of methane clathrate, we use the method developed by \citeA{Wakita:2022}.
As the strength of the material approaches zero at its melting temperature, we use a melting curve (i.e., dissociation curve) of methane clathrate that differs from that of water ice \cite{Sloan:2007,Levi:2014}. 
Experimental work shows that methane clathrate is 20--30 times stronger than water-ice at 250--287 K and 50--100 MPa \cite{Durham:2003}.
Extrapolation of these results to Titan-relevant conditions \cite{Wakita:2022} suggests that the strength of methane clathrate depends on temperature as less than the strength of pure water-ice, resulting in both materials having similar strengths below 100 K (Figure \ref{fig:tmp_yld}b).
Because there is no available shock physics equation of state (EOS) or Hugoniot data for methane clathrate, we use the analytical equation of state (ANEOS) of water-ice for both water-ice and methane clathrate, similar to previous work \cite{Wakita:2023}. 
For sufficiently large impactors, the presence of the ocean can influence the dynamics of crater formation. 
Previous work has shown that these effects become apparent for impactors larger than 5 km in diameter \cite{Wakita:2022}; for such impactors, we consider a subsurface ocean 100 km below the surface.
To achieve an appropriate density contrast between the ice and ocean, we use the simpler Tillotson EOS of water ice for icy materials \cite{Tillotson:1962,Ivanov:2002} and the ANEOS of water for the subsurface ocean \cite{Turtle:2001} for the $\geq 5$ km-diameter-impactors \cite<e.g.,>{Bray:2014,Silber:2017,Bjonnes:2022}. 
To evaluate the effects of this decision, we run simulations to test different EOSs for icy materials (Tillotson EOS and ANEOS) and different crustal structures (i.e., a 60 km thick ice shell without an ocean or a 100 km thick ice shell with an internal ocean; we consider a 5 km-diameter impactor and a 0--15 km thick methane-clathrate layer).
The resultant crater sizes are within $\sim$10\% for all EOSs and overall crustal structures (see Supplementary Text S1, Figure S1, Table S1, and S2). 
%0.12 for 0 km, 0.13 for 5km, 0.13 for 10 km, and 0.06 for 15 km
We confirm that these small variations affect our estimate of Titan's surface age by at most 1.8 \%(see Supplementary Text S1 and Figure S2).
We also note that we do not consider porosity within target, as there is no observational data on Titan's surface porosity. 
Porosity, however, would primarily affect smaller craters that have transient crater depths that are comparable to the thickness of the porous layer. 
For example, the transient crater depth of a 1.5 km-diameter-impactor is 5--7 km depending on the thickness of the methane-clathrate layer. 
Unless significant amounts of porosity exists at greater depths, larger craters should not be substantially affected by porosity.
This is supported by \cite{Kalousova:2024} who showed that the differences in diameter between craters formed in a non-porous methane-clathrate layer and those formed in porous crusts (10\% and 20\%) are less than 10\%, suggesting that the effect of porosity will be minor in these cases (a 4 km-diameter-impactor on a 10 km thick methane-clathrate layer).
Further investigation may be needed, but our assumption of no porosity should be reasonable for our estimates of Titan's surface age. 

Care must be taken to accurately measure the crater diameter during the simulations.
When the impactor diameter is $\lesssim 3$ km, the rim location is obvious throughout the simulation. 
For impactors $\gtrsim 3$ km in diameter, however, their transient craters become deeper where warmer and weaker material exists ($\gtrsim$ 15km depth from the surface, see Figure \ref{fig:tmp_yld}). 
As a result, overflow of warm material occurs at later times in the simulation and covers the pronounced rim \cite<see Figure 4 in>{Wakita:2023}.
Note that this occurrence is also observed in other cases of a higher temperature gradient at the surface \cite<e.g., Europa or Titan,>[see their Supplemental Movie]{Silber:2017,Wakita:2023}.
To determine the diameter in these scenarios, we idetnify the rim location when the rim is first established, i.e., between the crater collapse and the central uplift, following previous work \cite{Silber:2017,Wakita:2023}. 
Finally, impactors $\geq 20$ km in diameter produce huge central uplifts, whose collapse covers the crater rim before the rim is established, regardless of target structure. 
In such cases, it is impossible to identify the rim at the time of its formation and we, therefore, determine the crater rim at 6000 s after the impact, when the crater formation process (including collapse of the central uplift) has ceased. 

\section{Results} \label{sec:results}
We summarize our results quantifying the effects that methane clathrates have on crater diameter in Figure \ref{fig:dimp_dcrater}a and Tables S1 and S2. 
Note that oblique impacts occur more frequently than vertical impacts and the most probable impact angle is 45$^\circ$ \cite{Shoemaker:1962}. 
We note that while Titan's atmosphere will reduce the velocity of low-angle impacts more than vertical impacts, which will influence the resulting crater morphology, such effects are not expected to be significant for the sized impactors we consider.
Therefore, we scale our vertical impact results so that they represent oblique impacts at 45$^\circ$ following \citeA{Johnson:2016c}, i.e., multiplying $\sin^{0.38}(45^{\circ})$ (see Tables S1 and S2). 

\begin{figure}
\noindent\includegraphics[width=0.8\textwidth]{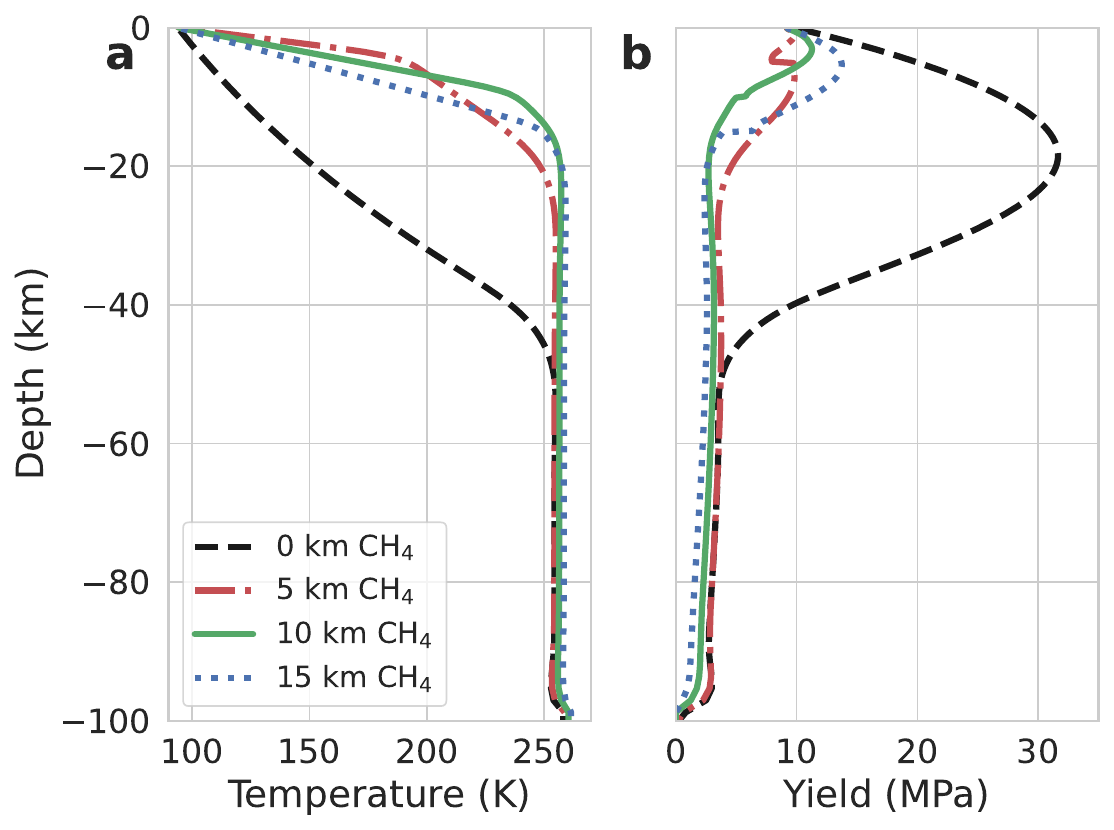}
\caption{Variations in temperature (a) and yield strength (b) with depth from Titan's surface.
Each line depicts a different methane-clathrate thickness (see legend).
The temperature profiles follow the 1 mm ice grain results of \citeA{Kalousova:2020}.
We model the yield strength based on temperature and other material parameters \cite<see>{Wakita:2022,Wakita:2023}. 
}
\label{fig:tmp_yld}
\end{figure}

We find that craters formed by impactors less than 2.5 km in diameter are relatively insensitive to the methane clathrate layer thickness. 
This is because the transient craters formed by these small impactors are shallow and contained entirely within the methane-clathrate layer ($<$ 15 km), so the strength difference between the methane-clathrate and water-ice crust has little effect on the crater diameter (see Figure \ref{fig:tmp_yld}b).
For larger impactors, the transient crater breaches the methane-clathrate layer, and as a result, the clathrate’s thermal properties dominate. 
Since methane clathrates are highly insulating, the temperatures at depths beneath the clathrate layers are higher than the temperatures at comparable depths in the case of pure water-ice (i.e., 0 km methane clathrate).
As such, the strength of the crust at these depths is weaker when a methane-clathrate layer exists (Figure \ref{fig:tmp_yld}b).
Craters that reach this warmer and weaker ice beneath the clathrate layer are $\sim$1.7 times larger in diameter compared to those formed in a pure water-ice target (Figure \ref{fig:dimp_dcrater}a).
The transition between impactors that form transient craters contained within the clathrate layer to those that breach it results in a distinct jump in crater size.
The impactor diameter at which this jump occurs depends on the strength within the ice crust (Figure \ref{fig:tmp_yld}b), which varies with the methane-clathrate layer thickness; this transition occurs at 2--3.5 km-diameter impactors for a 10 km thick methane-clathrate layer, and at 3.5--4 km diameter for 5 and 15 km thick clathrate layers.

When the impactor is larger than 20 km in diameter, the resulting crater diameter becomes relatively insensitive to the target structure (i.e., 390--520 km for 20 km-diameter-impactor and 492--680 km for 25 km, see Table S1).
This is because the yield strength of the pure water-ice target at the base of the transient crater ($\sim 80$ km) reaches a similar weak value to that observed in the warm ice between the methane-clathrate layer in those scenarios: (1.52--2.95 MPa, Figure \ref{fig:tmp_yld}b). 
This causes a jump in crater diameter at the 16--20 km impactor diameter range for the pure water-ice case (i.e., 0 km methane-clathrate layer, see also Figure \ref{fig:dimp_dcrater}a).

From our impact simulation results, we derive crater scaling laws for impacts on Titan. 
Based on the thickness of the methane-clathrate layer, we describe the relationships between impactor diameters ($d_{\rm imp}$) and crater diameters ($D_{\rm crater}$). 
The following equations are our crater scaling laws in units of km (see also lines in Figure \ref{fig:dimp_dcrater}b): \\
For a 0 km thick methane-clathrate layer, 
\begin{equation}
\log(D_{\rm crater}) = 
\begin{cases}
0.944 \log(d_{\rm imp}) + 1.160, & \text{if $d_{\rm imp} < 16.13$}, \\
2.484 \log(d_{\rm imp}) - 0.699, & \text{if $16.13 \leq d_{\rm imp} < 20.0$}, \\
1.041 \log(d_{\rm imp}) + 1.179, & \text{if $20.0 \leq d_{\rm imp}$}.
\end{cases}
\label{eq:law0}
\end{equation}
For a 5 km thick methane-clathrate layer, 
\begin{equation}
\log(D_{\rm crater}) = 
\begin{cases}
0.918 \log(d_{\rm imp}) + 1.225, & \text{if $d_{\rm imp} < 3.536$}, \\
3.645 \log(d_{\rm imp}) - 0.269, & \text{if $3.536 \leq d_{\rm imp} < 3.786$}, \\
1.075 \log(d_{\rm imp}) + 1.216, & \text{if $3.786 \leq d_{\rm imp}$}.
\end{cases}
\label{eq:law5}
\end{equation}
For a 10 km thick methane-clathrate layer, 
\begin{equation}
\log(D_{\rm crater}) = 
\begin{cases}
0.788 \log(d_{\rm imp}) + 1.246, & \text{if $d_{\rm imp} < 2.479$}, \\
2.310 \log(d_{\rm imp}) + 0.646, & \text{if $2.479 \leq d_{\rm imp} < 3.091$}, \\
1.071 \log(d_{\rm imp}) + 1.253, & \text{if $3.091 \leq d_{\rm imp}$}.
\end{cases}
\label{eq:law10}
\end{equation}
For a 15 km thick methane-clathrate layer, 
\begin{equation}
\log(D_{\rm crater}) = 
\begin{cases}
0.753 \log(d_{\rm imp}) + 1.234, & \text{if $d_{\rm imp} < 3.485$}, \\
3.825 \log(d_{\rm imp}) - 0.430, & \text{if $3.485 \leq d_{\rm imp} < 4.098$}, \\
1.048 \log(d_{\rm imp}) + 1.270, & \text{if $4.098 \leq d_{\rm imp}$}.
\end{cases}
\label{eq:law15}
\end{equation}
Note that we consider jumps as described in the previous section and assume a linear fit in a log-log space. 
Our crater scaling laws are relatively simple because there is no need to convert from the transient crater diameter to the final crater diameter, as in previous studies \cite{Artemieva:2005,Korycansky:2005}.
Although there are several jumps in our results, the overall scaling laws generally fall between previous crater scaling laws for water and sand targets \cite<see lines in Figure \ref{fig:dimp_dcrater}a,>{Artemieva:2005,Korycansky:2005}.
For $\leq 50$ km diameter craters, our scaling laws are relatively close to that for rock \cite{Johnson:2016c}, indicating that cold strong ice behaves more like rock than water or sand (see Figure \ref{fig:tmp_yld}).
For $\geq 50$ km craters, especially on the methane-clathrate layer (i.e., weaker target), our results are closer to the scaling law for water. 
These similarities with other scaling laws validate that ours with jumps well represent the change from stronger to weaker targets. 
In contrast, the sand-target crater scaling law underestimates the crater diameter of any size, suggesting that it might be a poor proxy for Titan. 
These will lead to similarity and discrepancy in estimating the surface age (see below).
We mention, however, that our results are only valid for impactors between 1.5--25 km with an impact velocity of 10.5 km/s at 45$^\circ$ (i.e., crater diameters between 22--600 km). 
Smaller impactors are likely to be influenced by passage through Titan's atmosphere \cite{Artemieva:2003}, and effects such as ablation would need to be considered before applying these crater scaling laws. 
Impactors traveling at higher/lower velocities than those considered here will result in larger/smaller craters than predicted by these crater scaling laws \cite<e.g., $v_{\rm imp}^{0.5}$, see>{Johnson:2016c}.

Finally, we estimate Titan's surface age using our new crater scaling laws. 
We use the heliocentric impactor rate on Titan assuming it has been constant over the age of Titan's surface \cite{Zahnle:2003}, and fit the results to the cumulative number of craters observed on Titan \cite{Hedgepeth:2020}.
Because our crater scaling laws are limited to craters larger than 22 km (see Figure \ref{fig:dimp_dcrater} and Table S1), we only use the number of $\geq 22$ km craters to compare to the observed data. 
We find that Titan's surface age is 300--340 Myr, depending on the thickness of the methane-clathrate layer (Figure \ref{fig:age}), or $\sim$420 Myr if no clathrates are present. 
Since the craters on the pure water-ice layer are smaller than those on the methane-clathrate layer (Figure \ref{fig:dimp_dcrater}), the surface age for a water-ice surface is about 100 Myr older. 
The estimated surface age differences among the 5--15 km thick methane-clathrate layer cases are minor (i.e., 300--340 Myr), because their crater scaling laws are similar (see Figure \ref{fig:dimp_dcrater}b and Equations (\ref{eq:law5})--(\ref{eq:law15})).

\section{Discussion} \label{sec:discussion}
\subsection{Implications of a younger surface age}
Our results suggest that Titan's surface age is 300--340 Myr (298$\pm$21.3--338$\pm$31.5 Myr, Figure \ref{fig:age}), assuming a surface methane-clathrate layer, and 420 Myr if no clathrates are present. 
This is similar to the younger age of the previous estimates based on the crater scaling law of water \cite<i.e., 100--200 Myr,>{Lorenz:2007,Neish:2012,Hedgepeth:2022}.
As mentioned above, this is likely due to the fact that, unlike colder pure water-ice, which more closely follows a crater scaling law of rock, the warmer temperature profiles of targets with methane clathrate surface layers result in craters that more closely follow the crater scaling laws of water\cite<see Figure \ref{fig:dimp_dcrater}a,>{Artemieva:2005}.
These estimates, however, rely on the assumed impactor rate on Titan, which has some uncertainty.
We use the heliocentric impact rate on Titan from \citeA{Zahnle:2003}, who proposed that it is $5.4 \times 10^{-5}$ the annual impact rate of comets on Jupiter (0.005 per year).
\citeA{Nesvorny:2023} reported a slightly lower rate ($4.5 \times 10^{-5}$), which considers the destruction of comets based on their physical lifetime \cite{Nesvorny:2017}. 
If we use the impact rate of \citeA{Nesvorny:2023}, Titan's surface age would be 1.2 times older (i.e., 360--500 Myr). 
We acknowledge that the impact rates have uncertainties \cite<i.e., 0.005$^{+0.006}_{-0.003}$ per year,>{Zahnle:2003}, because of an incomplete understanding of comets (e.g., number, mass, size).
Nevertheless, both of these impact rates rule out the older age estimates \cite<i.e., 1000 Myr, >{Neish:2012,Hedgepeth:2022}.

As mentioned in Section \ref{sec:introduction}, the age that we derive is most accurately referred to as the crater retention age; it indicates the duration over which a surface can retain observable craters (in this case, observable in Cassini remote sensing observations). 
A crater retention age may reflect the time since a resurfacing event (e.g., the eruption of the lunar maria).
Alternatively, it may reflect the balance between crater formation and destruction rates.
For example, Earth's crater retention age is a few hundred million years \cite{Kenkmann:2021}, which reflects erosion and burial processes, along with active tectonism. 
On Titan, several exogenic and endogenic processes can erase craters after they are formed, including fluvial erosion and infill, aeolian infill, viscoelastic relaxation, and resurfacing due to cryovolcanism \cite{Neish:2013,Neish:2016,Schurmeier:2024,Soderblom:2009,Soderblom:2010,Radebaugh:2011}.
Thus, our derived surface age of a few hundred million years can be used to infer the rates of different processes on Titan (i.e., erosion and/or relaxation).

Some of these crater-erasing processes correlate with the presence of methane in Titan's atmosphere, through the influence of its methane-based hydrological cycle. 
The lifetime of Titan's atmosphere has been estimated from a number of different sources, and ranges from 30 Myr to 300--500 Myr \cite{Horst:2017}. 
It would take $\sim$30 Myr to destroy the current methane in Titan's atmosphere due to photolysis \cite{Yung:1984,Wilson:2004}, but most ages from the models based on isotopes and/or ions overlap around 300--500 Myr \cite{Nixon:2012,Mandt:2012,Horst:2008}.
Intriguingly, these ages overlap with our estimated age based on impact craters over the methane-clathrate layer. 

There are several different ways of explaining this correlation. 
If methane outgassed from the interior $\sim$500 Myr ago \cite{Tobie:2006}, it may have resulted in a global surface methane--nitrogen ocean that could have efficiently erased existing craters and prevented the formation of observable craters in the marine environment.
Indeed, Cassini's observations revealed a dearth of craters in Titan's lowlands, consistent with this idea \cite{Neish:2014}. 
Although Cassini observations found methane lakes to be constrained to the polar regions \cite{Stofan:2007,Hayes:2008}, it is possible that a global methane ocean existed on Titan in the past \cite{Burr:2013,Larsson:2013}. 
The proposed timescale of the past ocean is 300--600 Myr \cite{Larsson:2013}, which is comparable to our estimates.
As such, the observable craters on Titan might have formed after the global ocean had disappeared or been reduced in size.
If the lowlands are merely liquid-saturated layers, and not completely submerged, that could also prevent the formation of observable craters \cite{Neish:2014,Wakita:2022}, but it may not be as effective at erasing pre-existing craters.

A large, catastrophic event could have also influenced Titan's surface a few hundred million years ago. 
There are several possibilities, including a global resurfacing event due to rapid climate change \cite{Moore:2012} and the formation of the Saturnian ring \cite{Wisdom:2022,Teodoro:2023}, or the formation of Titan itself \cite{Mosqueira:2003,Canup:2006,Anderson:2021,Sekine:2011,Asphaug:2013}. 
Although Titan's formation age is also uncertain, as well as its formation process, a relatively recent event could have erased previously existing geological features.
\citeA{Wisdom:2022} demonstrated that the destruction of the putative Chrysalis moon could have created Saturn's rings, and fragments of that body might have contributed to multiple impact events on Titan.
Note that they suggested that this destruction occurred around 100--200 Myr ago, close to our estimated surface age. 
\citeA{Teodoro:2023} also argued that the collisions between mid-sized moons caused the formation of Saturnian rings, which could provide additional debris that could impact Titan. 
If one or several large fragments were big enough to melt Titan's surface, it might globally refresh Titan's surface. 
These large impact events might have released a significant amount of methane from any clathrate layers there \cite{Wakita:2025a}, further altering Titan's climate and hence the exogenic processes occurring there.
We note, however, that the impact velocity of planetocentric impactors is about one-third that of heliocentric impactors \cite{Bell:2020}, resulting in the crater size being halved. 
If planetocentric impactors significantly contribute to Titan's current surface crater population, a revised analysis that includes the flux of these impactors would be required.

\subsection{The possible presence of a methane-clathrate layer}
We suggest that Titan's current crater size-frequency distribution might indicate the presence of a methane-clathrate layer, consistent with previous work \cite<e.g.,>{Schurmeier:2024}. 
The observed cumulative crater count shows a change in slope for mid-sized craters of 45.25--90.50 km in diameter (i.e., 32$\sqrt{2}$--64$\sqrt{2}$ km, see Figures \ref{fig:age}b and S3a), consistent with a change in slope for our methane clathrate models that relates to the jump in crater diameter when transient craters breach the clathrate layer. 
To assess the presence of a slope change in the crater data, we conduct a Monte Carlo simulation based on the observed data by considering the crater diameters with their errors (Supplementary Text S2).
We confirm the slope of -0.18 for the 32$\sqrt{2}$--64$\sqrt{2}$ km crater range is noticeably shallower than the slope for the larger or smaller crater size ranges (see Supplemental Text S2 and Figure S3B). 
Additionally, we compare observed ‘binned’ crater count with a ‘raw’ cumulative crater count (Figure S4). 
Note that the ‘raw’ count is not the exact raw count but is corrected based on the probability of detection \cite{Hedgepeth:2020}.
The ‘raw’ count also exhibits a clear slope change.
Using our crater scaling laws, we also found a methane-clathrate layer better represents the change in slope in the cumulative crater count than a 0 km thick layer (i.e., pure water-ice layer).
In addition to reproducing this change in slope, the 5 km case represents the best-fit to the cumulative crater count (Figure \ref{fig:age}b). 
Although there are several assumptions in deriving the crater scaling laws to illustrate those slopes (such as 45$^{\circ}$ impacts at 10.5 km/s), the presence of a methane-clathrate layer may better explain the present crater distribution on Titan.
If methane raining onto the surface results in the formation of a methane-clathrate layer, our surface age could therefore indicate a minimum age for the methane cycle on Titan. 

It is also important to note, however, that other conditions can accomplish the temperature profile of a methane-clathrate layer. 
\citeA{Downey:2025} find that a surface heat flux of 25-km-thick fully conductive water-ice crust could be as high as 42 mW/m$^2$ due to tidal heating. 
A 42 mW/m$^2$ surface heat flux would result in a thermal gradient that is three times steeper than for our pure water-ice case \cite{Kalousova:2020}, and is most similar to our 10-km-thick methane-clathrate-layer case. 
The scaling law for such a target would approach those for our methane-clathrate cases (Figure \ref{fig:dimp_dcrater}b).
If surface methane-clathrate is included with tidal heating, the convective temperature increases \cite{Kalousova:2020}, which should lead to a thinner stagnant lid and a decrease in the diameter of the jump observed in our scaling laws.
Alternatively, an organic layer could have a lower thermal conductivity than that of methane clathrate, which would lead to enhanced subsurface temperatures \cite{Schurmeier:2018}. 
Nonetheless, if such a layer is thin (i.e., $\sim$100 m), its temperature gradient won’t be comparable to that of a thick ($>$5 km) methane clathrate layer.

\section{Conclusions} \label{sec:conclusions}
\begin{figure}
\noindent\includegraphics[width=1.\textwidth]{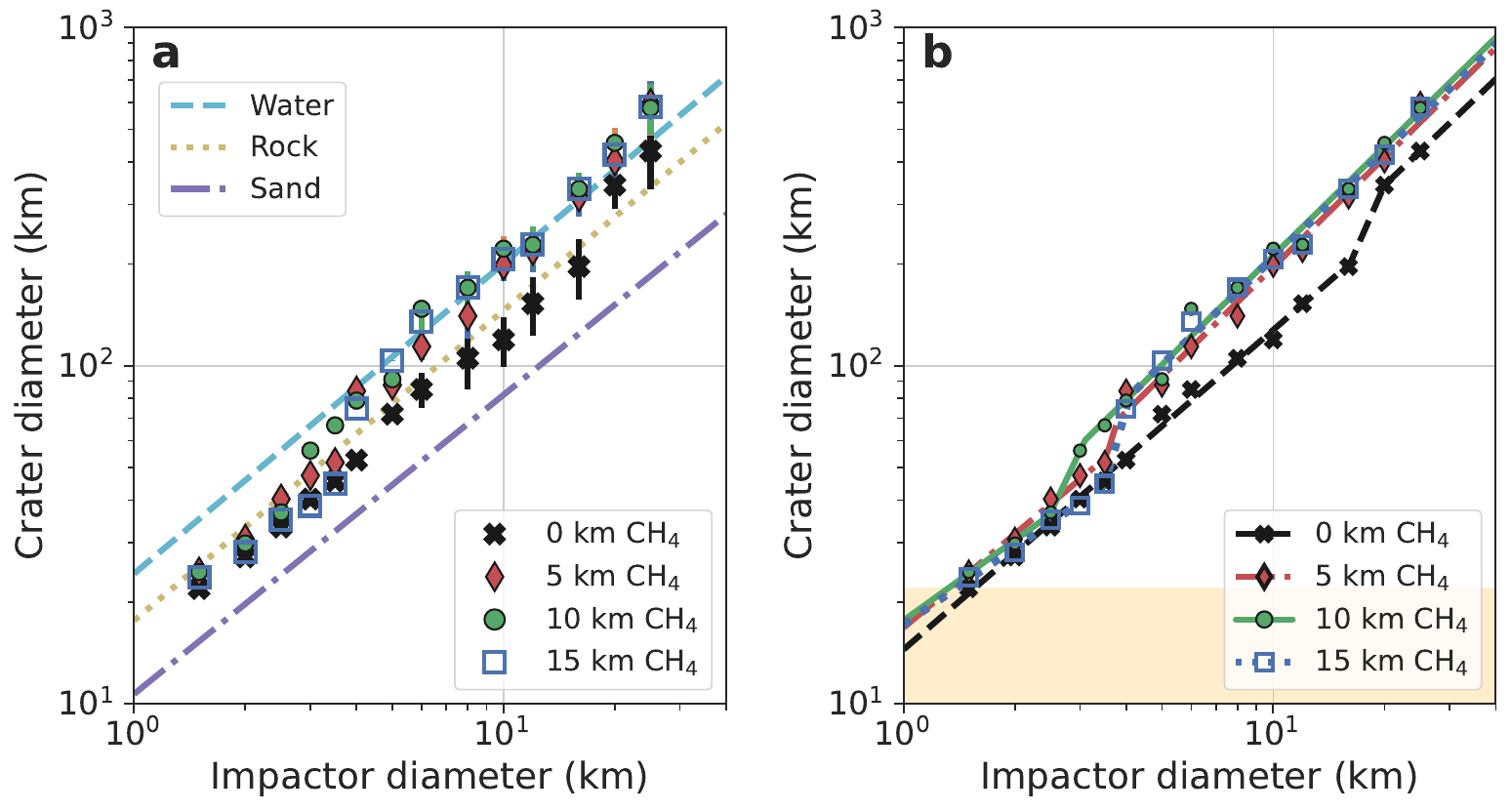}
\caption{Crater diameter as a function of impactor diameter.
Each color depicts a different methane-clathrate thickness (see legend).
Symbols in panel a represent our impact simulation results with error bars (see Tables S1 and S2). 
Also shown on Panel a are published crater scaling laws for water \cite{Artemieva:2005}, rock \cite{Johnson:2016c}, and sand \cite{Korycansky:2005}.
Panel b indicates our crater scaling laws for different methane-clathrate layer thickness (see Equations (\ref{eq:law0})--(\ref{eq:law15})).
The shaded region indicates craters less than 22 km in diameter, which are influenced by Titan's atmosphere \cite<e.g.,>{Artemieva:2003}. 
}
\label{fig:dimp_dcrater}
\end{figure}

\begin{figure}
\noindent\includegraphics[width=1.\textwidth]{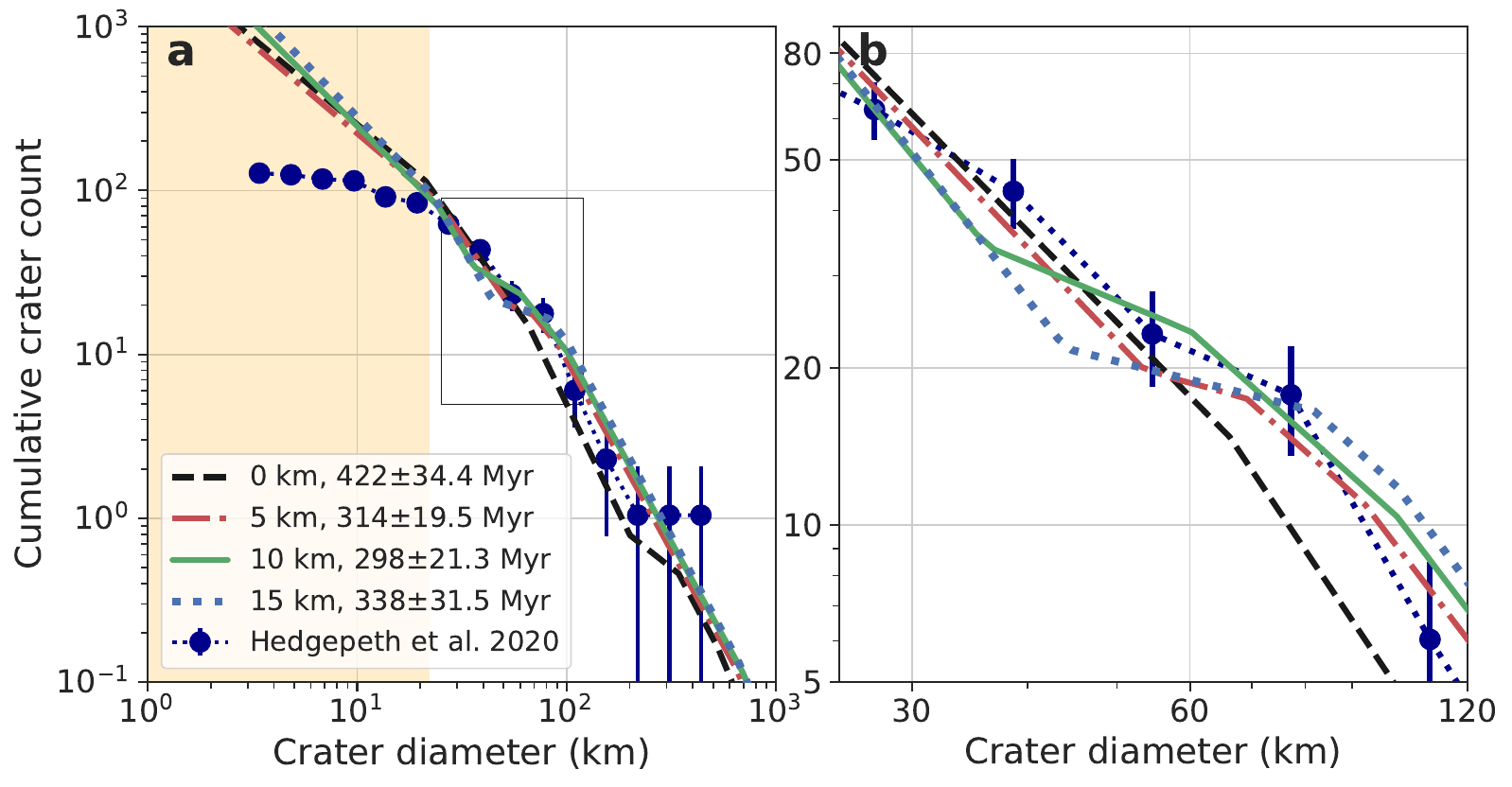}
\caption{Cumulative crater count as a function of crater diameter.
Each symbol and line depicts a fit to a different methane-clathrate thickness (see legend).
The darkblue dotted line shows Titan's observed crater count \cite{Hedgepeth:2020}.
The shaded region indicates craters less than 22 km, which is the lower limit of our results. 
Note that those craters are affected by Titan's atmosphere, hence the drop off from the trend line.
Panel b shows the inset of Panel a (a black box), so the details near the change in slope can be seen more clearly. 
}
\label{fig:age}
\end{figure}

We simulate impacts into a Titan-like target with and without a methane-clathrate layer to revise its surface age. 
We have derived new crater scaling laws according to the thickness of the methane-clathrate layer. 
Impact craters formed in a target with a surface layer of methane clathrates are about 1.7 times larger than those formed in a pure water-ice target, when the impactor diameter is between 3 km and 16 km (see Figure \ref{fig:dimp_dcrater}a).  
While these scaling laws are unique to Titan, our results highlight the importance of clathrates on crater formation and indicate that their insulating effects should be considered when studying impacts on any world with significant clathrate deposits.
On Titan, this effect results in a younger estimated surface age if a surface methane-clathrate layer is included in the model, as compared to a pure water ice target. 
Our results indicate that the size-frequency distribution of craters on Titan is more consistent with a surface methane-clathrate layer, supporting this younger age.
Our revised estimate of Titan's surface age is 300--340 Myr, assuming the presence of a surface methane-clathrate layer, which is a narrower range compared to the previous estimate of 200--1000 Myr \cite{Neish:2012,Hedgepeth:2020}. 
Our findings provide insights into the landscape evolution of Titan, and may be consistent with the inferred lifetime of Titan's methane-rich atmosphere \cite{Horst:2017}.
Alternatively, our results may imply that a past global ocean or a global resurfacing event occurred a few hundred million years ago on Titan, potentially due to global climate change, or multiple large impacts associated with the Saturnian ring formation \cite<e.g.,>{Moore:2012,Neish:2014,Wisdom:2022,Teodoro:2023}. 

%%

%  Numbered lines in equations:
%  To add line numbers to lines in equations,
%  \begin{linenomath*}
%  \begin{equation}
%  \end{equation}
%  \end{linenomath*}

%% Enter Figures and Tables near as possible to where they are first mentioned:
%
% DO NOT USE \psfrag or \subfigure commands.
%
% Figure captions go below the figure.
% Table titles go above tables;  other caption information
%  should be placed in last line of the table, using
% \multicolumn2l{$^a$ This is a table note.}
%
%----------------
%
\clearpage

\clearpage

\section{Open Research}
All of our results were produced using iSALE-2D and our input and output files will be available in \citeA{Wakita:2025}. 
Please note that iSALE-2D is distributed on a case-by-case basis to academic users in the impact community.
It requires registration from the iSALE webpage (https://isale-code.github.io/); computational requirements are listed there.  
%%%%%%%%%%%%%%%%%%%%%%%%%%%%%%%%%%%%%%%%%%%%%%%

\acknowledgments
This work was supported by NASA Cassini Data Analysis Program grant 80NSSC23K0218.
We gratefully acknowledge the developers of iSALE-2D, including Gareth Collins, Kai W\"{u}nnemann, Dirk Elbeshausen, Tom Davison, Boris Ivanov, and Jay Melosh. 
This research was supported in part through computational resources provided by Information Technology at Purdue, West Lafayette, Indiana. 

\section*{Conflict of Interest}
The authors declare no conflicts of interest relevant to this study.

%% ------------------------------------------------------------------------ %%
%% References and Citations

%%%%%%%%%%%%%%%%%%%%%%%%%%%%%%%%%%%%%%%%%%%%%%%
%
% \bibliography{<name of your .bib file>} don't specify the file extension
%
% don't specify bibliographystyle

% In the References section, cite the data/software described in the Availability Statement (this includes primary and processed data used for your research). For details on data/software citation as well as examples, see the Data & Software Citation section of the Data & Software for Authors guidance
% https://www.agu.org/Publish-with-AGU/Publish/Author-Resources/Data-and-Software-for-Authors#citation

%%%%%%%%%%%%%%%%%%%%%%%%%%%%%%%%%%%%%%%%%%%%%%%

%\bibliography{references.bib}

%Reference citation instructions and examples:
%
% Please use ONLY \cite and \citeA for reference citations.
% \cite for parenthetical references
% ...as shown in recent studies (Simpson et al., 2019)
% \citeA for in-text citations
% ...Simpson et al. (2019) have shown...
%
%
%...as shown by \citeA{jskilby}.
%...as shown by \citeA{lewin76}, \citeA{carson86}, \citeA{bartoldy02}, and \citeA{rinaldi03}.
%...has been shown \cite{jskilbye}.
%...has been shown \cite{lewin76,carson86,bartoldy02,rinaldi03}.
%... \cite <i.e.>[]{lewin76,carson86,bartoldy02,rinaldi03}.
%...has been shown by \cite <e.g.,>[and others]{lewin76}.
%
% apacite uses < > for prenotes and [ ] for postnotes
% DO NOT use other cite commands (e.g., \citet, \citep, \citeyear, \citealp, etc.).
% \nocite is okay to use to add references from your Supporting Information
%

\end{document}